\documentclass[%
 aip,
 amsmath,amssymb,
 reprint,%
]{revtex4-1}

\usepackage{graphicx}% Include figure files
\usepackage{dcolumn}% Align table columns on decimal point
\usepackage{bm}% bold math
\usepackage[inkscapelatex=false]{svg}
\usepackage[utf8]{inputenc}
\usepackage[T1]{fontenc}
\usepackage{mathptmx}
\usepackage{etoolbox}
\usepackage{soul}

\makeatletter
\def\@email#1#2{%
 \endgroup
 \patchcmd{\titleblock@produce}
  {\frontmatter@RRAPformat}
  {\frontmatter@RRAPformat{\produce@RRAP{*#1\href{mailto:#2}{#2}}}\frontmatter@RRAPformat}
  {}{}
}%
\makeatother

\begin{document}

\preprint{AIP/123-QED}

\title{Epitaxial high-K barrier AlBN/GaN HEMTs}
%\title{AlBN: Integrating High-K Dielectric Functionality in GaN High Electron Mobility Transistors}

\author{\fontsize{13}{13}Chandrashekhar Savant\textsuperscript{1}$^\dagger$, Thai-Son Nguyen\textsuperscript{1}$^\dagger$, Kazuki Nomoto\textsuperscript{2}$^\dagger$, Saurabh Vishwakarma\textsuperscript{3}, Siyuan Ma\textsuperscript{1}, Akshey Dhar\textsuperscript{1}, Yu-Hsin Chen\textsuperscript{1}, Joseph Casamento\textsuperscript{4}, David J. Smith\textsuperscript{5}, Huili Grace Xing\textsuperscript{1,2,6}, and Debdeep Jena\textsuperscript{1,2,6,7}\\
\fontsize{10}{11}\selectfont \textsuperscript{1}Department of Materials Science and Engineering, Cornell University, Ithaca, NY 14853, USA, \\ \textsuperscript{2}School of Electrical and Computer Engineering, Cornell University, Ithaca, NY 14853, USA, 
\\ \textsuperscript{3}School for Engineering of Matter, Transport and Energy, Arizona State University, Tempe, AZ 85287, USA, \\\textsuperscript{4}Department of Materials Science and Engineering, Massachusetts Institute of Technology, Cambridge, Massachusetts 02139, USA, \\ \textsuperscript{5}Department of Physics, Arizona State University, Tempe, AZ 85287, USA, \\\textsuperscript{6}Kavli Institute at Cornell for Nanoscale Science, Cornell University, Ithaca, NY 14853, USA, \\\textsuperscript{7}School of Applied and Engineering Physics, Cornell University, Ithaca, NY 14853, USA 
\\($^\dagger$Equal Contribution. Email: cps259@cornell.edu)
}

\begin{abstract}
We report a polarization-induced 2D electron gas (2DEG) at an epitaxial AlBN/GaN heterojunction grown on a SiC substrate.  Using this 2DEG in a long conducting channel, we realize ultra-thin barrier AlBN/GaN high electron mobility transistors that exhibit current densities > 0.25 A/mm, clean current saturation, a low pinch-off voltage of -0.43 V, and a peak transconductance of $\sim$0.14 S/mm. Transistor performance in this preliminary realization is limited by the contact resistance. Capacitance-voltage measurements reveal that introducing $\sim$7 \% B in the epitaxial AlBN barrier on GaN boosts the relative dielectric constant of AlBN to $\epsilon_{\rm r}^{\rm AlBN} \sim16$, higher than the AlN dielectric constant of $\epsilon_{\rm r}^{\rm AlN} \sim9$.  Epitaxial high-K barrier AlBN/GaN HEMTs can thus extend performance beyond the capabilities of current GaN transistors.
\end{abstract}

\maketitle

% \begin{quotation}

% \end{quotation}

%--------------------------start text-----------------------

% ------------------Introduction---------------

High electron mobility transistors (HEMTs) using the semiconductor GaN achieve high-voltage and high-frequency operation due to the wide bandgap and polarization fields in epitaxial heterostructures that give high electron velocities, breakdown fields, mobile carrier densities, and electron mobility.\cite{khan1992observation, ambacher1999two, asif1993high, mishra2002algan} The 2D electron gas (2DEG) channel induced at the AlGaN/GaN heterojunction by polarization discontinuity offers six times higher electron mobility than Si MOSFETs.\cite{ambacher1999two, mishra2002algan, thompson_90-nm_2004} GaN HEMTs offer a compelling energy efficiency advantage in computation by conditioning and delivering power to silicon microprocessors via heterogeneous integration in a manner unmatched by any other semiconductor.{\cite{Intel2023IEDM_DrGaN} GaN HEMT RF power amplifiers add communication capability to silicon.{\cite{IntelThen2019IEDMGaNonSi, intelMIT2020_GaNlogiconSi, IntelThen2021GaNon300mmSi, IntelThen2022_GaNon300mm111Si} Today's state-of-the-art HEMTs for power electronics and RF applications utilize AlGaN barrier layers on a GaN channel grown epitaxially on silicon, SiC, or sapphire substrates.\cite{pengelly_review_2012, tsao_ultrawidebandgap_2018, rosker2009darpa}

Developed in the early 2000s and deployed in the past decade, such conventional AlGaN/GaN heterostructures are approaching their performance limits in HEMTs.\cite{islam2022reliability, pengelly_review_2012} The dielectric constant $\epsilon_r$ of typical AlGaN barriers with Al mole fraction $\sim0.3$ is $\sim$ 8.4 - 9 and the energy bandgap is $\sim$4.2 eV.\cite{Jena2022quantum} The modest dielectric constant, energy bandgap (and band offset), and tensile strain present in AlGaN now limit the HEMT performance. Similar limitations of barrier properties of SiO$_2$ in Si CMOS in the early 2000s were solved by introducing high-K dielectric HfO$_2$ barriers in 2006, enhancing device performance and enabling Moore's law scaling.\cite{salahuddin_era_2018} In HfO$_2$, ferroelectricity was later discovered by doping with Zr, Si, or La, enabling FerroFET memories as a bonus.\cite{salahuddin_era_2018, boscke_ferroelectricity_2011, savant_dopant_2022} Nitride HEMTs now need hi-K barriers analogous to Hf$_{0.5}$Zr$_{0.5}$O$_2$ (HZO) for silicon, but which must simultaneously be ultrawide bandgap and epitaxial to provide the polarization-induced conductive channels.

Investigation of alternate barriers for GaN HEMTs, such as lattice-matched AlInN and binary AlN, have yielded encouraging results, but the dielectric constant of these barriers is still modest.  Over the last few years, AlScN has emerged as an interesting option, thanks to the discovery of ferroelectricity\cite{fichtner2019alscn}, high-K dielectric properties\cite{casamento2022highk}, and lattice-matching to GaN\cite{van2023dbr, TS2024Multilayer, dinh2023lattice}.  Although AlScN barrier RF and mm-wave GaN HEMTs\cite{green_scalngan_2019, kazior_high_2019, casamento2022transport, nomoto2025alscn, AlScNIEDM2024} and FerroHEMTs\cite{casamento2022ferrohemts, casamento2023review} have been demonstrated, the semiconductor community is addressing challenges of chemical oxidation \cite{casamento2020oxygen, greenwood_chemistry_1997}, energy bandgap reduction\cite{baeumler2019AlScNBG}, structural instabilities and phase separation in high Sc-containing AlScN.

Ferroelectric {\cite{Hayden2021AlBN, calderon2023atomic, savant2024APLFerro, savant2024PSSRRL} and high-K dielectric properties \cite{Hayden2021AlBN} were recently reported in AlBN layers.  The large energy bandgap and chemical inertness make AlBN barriers worth investigating as an alternate barrier material for high-voltage, high-temperature GaN transistors.  Here, we report the epitaxial integration of the AlBN gate barrier on the GaN platform and the realization of AlBN/GaN HEMTs.  We grew AlBN/GaN heterostructures by MBE on 3-inch diameter 6H-SiC wafers and observed polarization-induced 2DEGs.  We then fabricated AlBN/GaN HEMTs using alloyed metallic Ta/Al/Ni/Au ohmic `claw' contacts and standard Schottky gates and observed encouraging transistor characteristics. We find that the addition of $\sim$7 \% B boosts the dielectric constant of AlN by nearly two times to $\epsilon_{\rm r}^{\rm AlBN}$ = 16 at 2 MHz at 20 $^\circ$C.  By measuring the capacitance-voltage characteristics of thin AlBN barrier/GaN heterostructure-based MIS capacitors in several devices and at various frequencies, we confirmed that epitaxial AlBN on GaN is a robust, hi-K epitaxial dielectric that simultaneously provides polarization-induced conductive channels.

%--------Figure 1---------------------------
\begin{figure*}
\includegraphics[width=\textwidth]{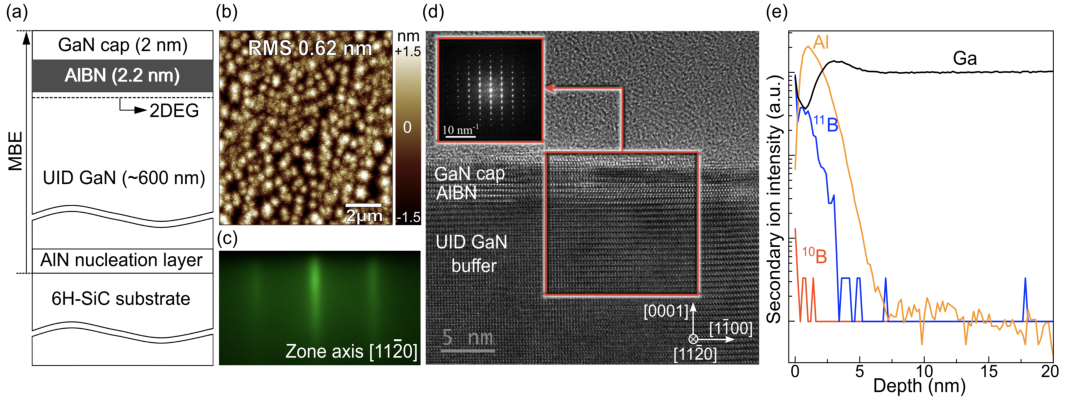}
\caption{\label{fig1} (a) Epitaxial AlBN/GaN heterostructure indicating a polarization-induced 2DEG at the heterojunction, (b) AFM micrograph, (c) RHEED pattern along [11-20] zone axis, (d) HRTEM micrograph of AlBN/GaN heterostructure.  Inset shows FFT analysis of red marked region indicating wurtzite phase.  (e) Composition-depth profiles of Al, Ga, $^{11}$B, and $^{10}$B species as measured by ToF-SIMS.}
\end{figure*}
%--------Figure 1---------------------------

% ------------------MBE growth, Film characterization------------------

%--------Figure 2---------------------------
%--------------------------
\begin{figure}[t!]
\centerline{\includegraphics[width=1\columnwidth]{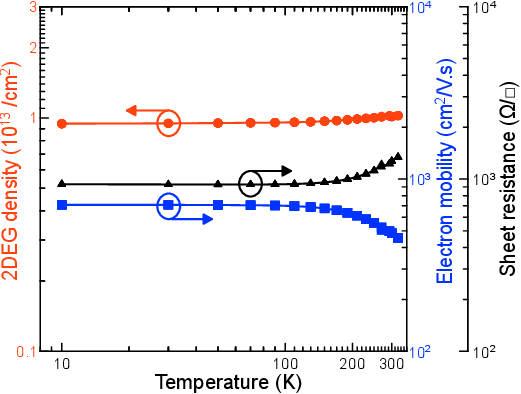}}
\caption{\label{fig2} Temperature-dependent Hall effect measurement data showing electron density, mobility, and sheet resistance from 320 K to 10 K confirming 2DEG properties in the AlBN/GaN heterostructure.  The measurement was performed with indium dots on the center 1 cm $\times$ 1 cm sample diced from the 3-inch diameter wafer.}
\end{figure}
%--------------------------
%--------Figure 2---------------------------

%--------Figure 3---------------------------
%-------------------------
\begin{figure*}
\includegraphics[width=\textwidth]{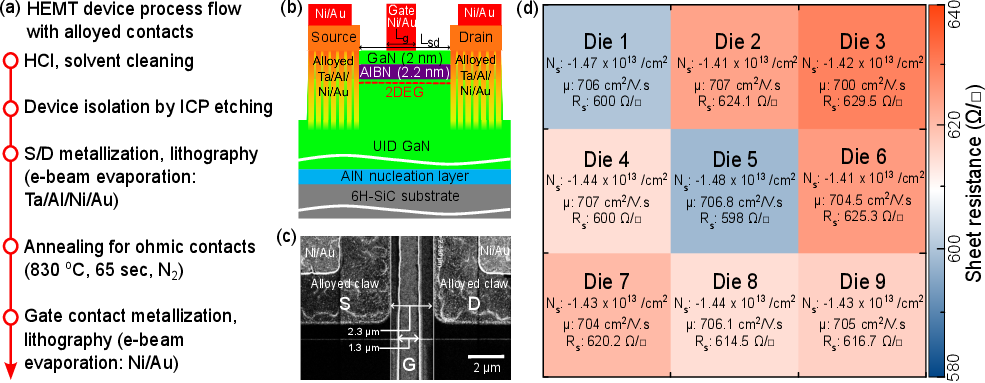}
\caption{\label{fig3} (a) The device process flow of group III-Nitride HEMTs with alloyed contacts, (b) Schematic cross-section of $\sim$7\% B containing AlBN barrier GaN HEMT showing the alloyed contacts, gate, and channel regions. (c) Scanning electron microscope (SEM) image of a processed HEMT showing source (S), drain (D), and gate (G) regions.  (d) Room temperature Hall-effect data for $\sim$7\% B containing AlBN barrier GaN HEMT measured after device processing.}
\end{figure*}
%--------------------------
%--------Figure 3---------------------------

Figure \ref{fig1}(a) schematically shows the AlBN/GaN HEMT heterostructure realized by MBE for this study.  We first grew a $\sim$125-nm AlN nucleation layer on the SiC substrate, starting with a nitrogen-rich AlN layer with III/V ratio $\sim$ 0.7 to capture surface impurities and block Si diffusion, and then transitioned to metal-rich AlN with III/V $\sim$ 1.1  to promote smooth AlN growth.\cite{hoke2006AlNonSiC} We then grew a 600-nm-thick unintentionally doped (UID) GaN layer under metal-rich conditions with a III/V ratio > 1, followed by $\sim$ 2.2-nm epitaxial wurtzite phase AlBN barrier film at a III/V flux ratio of 0.7. Nitrogen-rich growth conditions are necessary to grow AlBN because N has a thermodynamic preference to bond to Al rather than to B.\cite{hoke2007thermodynamic} We then deposited a 2-nm UID GaN cap to protect the barrier layer from surface contamination.

Figure \ref{fig1}(b) shows an AFM micrograph of a dislocation-mediated, yet smooth surface morphology of an as-grown AlBN/GaN heterostructure with 0.62-nm rms roughness over a 10 $\times$ 10 $\mu$m$^2$ scan area.  Figure \ref{fig1}(c) shows the streaky RHEED pattern of AlBN films along the [11-20] zone axis observed during MBE growth, confirming the AlBN film is epitaxial and of wurtzite phase.  Figure \ref{fig1}(d) shows a cross-sectional HRTEM image of the GaN/AlBN/UID GaN heterostructure up to $\sim40$-nm depth from the surface.  AlBN is in lighter contrast, sandwiched between the darker GaN cap and the UID GaN layers.  The micrograph shows the AlBN and GaN layers are crystalline. The AlBN/GaN interfaces are not as smooth as typical AlN/GaN interfaces, but the thin AlBN layer is continuous with slightly varying thickness.  The Fast Fourier transform (FFT) pattern of the region marked by the square red box in the inset of Figure \ref{fig1}(d) confirms the wurtzite phase of AlBN, GaN, and the epitaxial relationship between the layers.  The smearing of low-frequency FFT spots due to the slight difference in lattice parameters between GaN and AlBN also confirms the presence of the AlBN layer.  ToF-SIMS composition depth profile of $^{11}$B, Al, and Ga species in Figure \ref{fig1}(e) verifies the presence of boron in the AlBN layer.  We determined the B/(Al+B) ratio in the AlBN barrier layer to be 7.4\% by XPS measurements.  XRR measurements confirm $\sim$2-nm thickness for the GaN cap, $\sim$2.2-nm thickness for the AlBN barrier. TEM micrographs of the AlBN barrier sample further confirm the film thicknesses of the GaN cap to be 2-nm and the AlBN layer to be 2.2-nm.

%\textcolor{red}{and $\sim$2-nm for the AlN barrier [***which AlN barrier***???]}

Figure \ref{fig2} shows the results of temperature-dependent Hall effect measurements performed on diced 1 x 1 cm$^2$ samples from the center region of the as-grown 3-inch diameter AlBN/GaN heterostructure wafer using a van der Pauw geometry with indium contacts. We obtained this data by varying the temperature from 10 K to 320 K in a Lakeshore Hall measurement system at 1 T.  At 300 K, we observe a polarization-induced 2DEG sheet concentration $N_{\rm s}=$9.25 × 10$^{12}$ /cm$^{2}$, electron mobility $\mu=$524 cm$^{2}$/V·s, and a sheet resistance $R_{\rm s} =$1290 $\Omega$/sq in the AlBN/GaN heterostructure.  We observe < 6 \% decrease in the sheet density as temperature decreases from 320 K to 10 K, indicating minimal carrier freeze-out.  The low-temperature plateauing of the sheet carrier density, electron mobility, and sheet resistance confirm the presence of a typical 2DEG.\cite{casamento2022transport}  We observe the 2DEG across the large 3-inch wafer, with sheet density $N_{\rm s}$ between 7.1 - 9.3 $\times$ 10$^{12}$/cm$^{2}$. We re-measured the Hall data on an as-grown sample 16 months after the MBE growth to find a nearly unchanged sheet density, mobility, and sheet resistivity. The 2DEG density we measure in the 2.2-nm AlBN/GaN heterostructure is lower than the sheet density of 2.08 $\times$ 10$^{13}$/cm$^{2}$ in a comparable 2-nm AlN/GaN heterostructure\cite{casamento2022ferrohemts} due to the higher dielectric constant of AlBN relative to AlN.

% ------------------Device Process Flow, characteristics-----------------

%--------Figure 4---------------------------
%--------------------------
\begin{figure*}
\includegraphics[width=\textwidth]{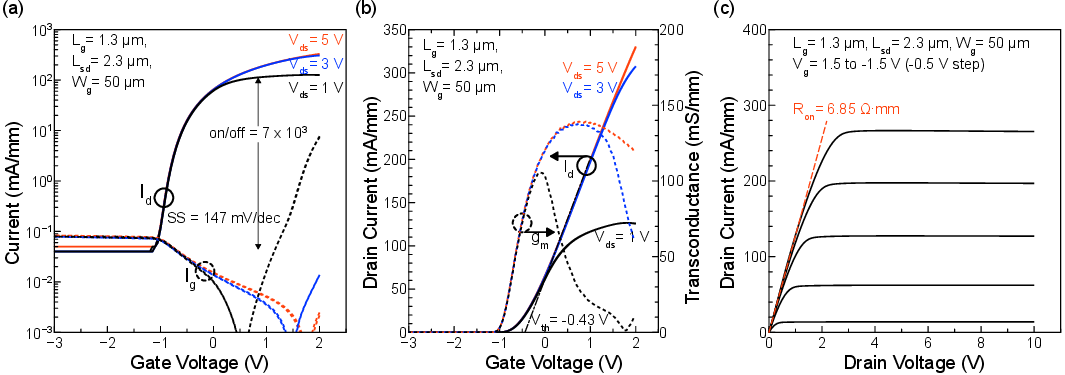}
\caption{\label{fig4} (a) Measured characteristics of the GaN HEMT with AlBN barrier with $\sim$7\% B: (a) log scale, and (b) and linear scale Transfer characteristics showing an on/off ratio exceeding three orders and Transconductance vs. $V_{\rm gs}$ showing a peak transconductance of 139 mS/mm and. (d) Output characteristics showing on resistance $R_{\rm on}= 6.8$ $\Omega$.mm with repeatable current saturation and a maximum drain current $I_{\rm d}$ of 280 mA/mm at a gate voltage $V_{\rm gs}=1.5$ V.
}
\end{figure*}
%--------------------------
%--------Figure 4---------------------------

In Figure \ref{fig3}(a), we have outlined the process flow used to realize the HEMTs shown in Figures \ref{fig3}(b) and (c) from 1 $\times$ 1 cm$^2$ samples diced from the center of the 3-inch diameter wafer.  Dry BCl$_3$ inductively-coupled plasma (ICP) was performed for device isolation and to simultaneously create trenches for metal sidewall contacts to the 2DEG.  We defined the source/drain regions with a SiO$_2$/Cr hard mask and deposited Ta/Al/Ni/Au (20/150/50/50-nm) metal stack\cite{lu2022lowalloy, malmros2011alloyedcontact} using e-beam evaporation for source/drain contacts, and annealed the stack at 830 $^\circ$C for 65 seconds in N$_2$ ambient.  We deposited a 30/220-nm Ni/Au gate metal stack directly on the sample surface and defined the gate lengths $L_{\rm g}$ by photolithography.  Figure \ref{fig3}(b) shows the AlBN/GaN HEMT device cross-section schematic with the alloyed claw source/drain contacts to the 2DEG channel.  Figure \ref{fig3}(c) shows the SEM image of a fully processed HEMT viewed from the top.  All HEMTs reported in this study have a source-to-drain spacing $L_{\rm sd}=$2.3 $\mu$m, device-width $W=$50 $\mu$m, and a rectangular gate placed in the middle of the source-to-drain spacing with a gate length $L_{\rm g}=$1.3 $\mu$m.

We repeated 300 K Hall effect measurements on processed van der Pauw patterns after device fabrication.  As shown in Figure \ref{fig3}(d), we measured 2DEG sheet densities $N_{\rm s}$ = 1.41 to 1.48 $\times$ 10$^{13}$ /cm$^2$, mobilities $\mu$ = 700 to 707 cm$^2$/V.s, and sheet resistances $R_{\rm s}$ = 598 to 630 $\Omega$/sq across different dies with alloyed contacts.  The higher mobility and sheet density and consequently lower sheet resistance after device processing compared to the as-grown data, presumably due to annealing steps during the device processing, need to be understood better in the future\cite{ben2018AlNannealing, biswas2015GaNAnnealing, chan20162DEGmobilityrecovery}. 

We assessed buffer leakage through the UID GaN layer by removing the top 60-nm of the MBE-grown AlBN/GaN heterostructure to remove the active region of some HEMTs by BCl$_3$ ICP etching, and found low leakage currents of $\sim$0.1 nA/mm at 20 V for 5-$\mu$m spacing between two alloyed contacts indicating its high resistivity. Transfer length measurements on the AlBN/GaN heterostructure devices revealed nonlinear IV characteristics with Schottky behavior of unoptimized alloyed contacts. The measured contact resistance $R_{\rm c}$ is $\sim$$1.44$ $\Omega \cdot$mm and drops to below $1$ $\Omega \cdot$mm when the current exceeds 150 mA/mm. The specific contact resistance is $\sim$5 $\times$ 10$^{-6}$ $\Omega \cdot$cm$^2$ and approaches $\sim$3.6 $\times$ 10$^{-6}$ $\Omega \cdot$cm$^2$ at 150 mA/mm.

Figures \ref{fig4}(a) and (b) show the log and linear scale HEMT transfer characteristics revealing $I_{\rm on}/I_{\rm off} \sim 7 \times 10^3$ limited by gate leakage.  We observe sharp pinch-off characteristics with a low threshold voltage $V_{\rm th}=$-0.43 V at different drain biases $V_{\rm ds}= $ 1, 3, and 5 V, suggesting minimal drain-induced barrier lowering (DIBL), with a minimum sub-threshold slope (SS) of 147 mV/decade and peak extrinsic transconductance $g_{\rm m}^{\rm ext}=$139 mS.mm$^{-1}$.  Figure \ref{fig4}(c) shows the HEMT output characteristics with clean current saturation of the drain current $I_{\rm d}$, its control with gate bias $V_{\rm gs}$ from on-state to pinch-off.  We measure a transistor on-resistance $R_{\rm on}=$6.85 $\Omega$.mm, and a saturation drain current $I_{\rm d}^{\rm sat}=$266 mA/mm at $V_{\rm gs}=$1.5 V. The transistor characteristics, currently limited by contact resistance, exhibit a low threshold voltage.  

We measured capacitance-voltage (C-V) characteristics on metal-insulator-semiconductor (MIS) capacitors of GaN-cap/AlBN/GaN structure using circular Ni/Au pads with 20, 40, and 80 $\mu$m diameters as top electrodes and the 2DEG accessed using alloyed contacts as the bottom electrode. Figure \ref{fig5} shows the measured C-V characteristics of the 2-nm GaN-cap/2.2-nm AlBN/GaN heterostructure-based MIS capacitor at 2 MHz AC at 20 $^\circ$C. The solid line shows the calculated C-V characteristics using a self-consistent 1D Schrödinger Poisson solution.\cite{gregsnider1DP} Parameters used in the simulation are the relative dielectric constant of GaN $\epsilon_{\rm r}^{\rm GaN}=10.4$, a polarization discontinuity of 11.8 $\mu$C/cm$^2$,\cite{liu2017wurtzite} at the AlBN/GaN heterojunction, an AlBN bandgap of $E_{\rm g}^{\rm AlBN}$ = 6.1 eV and an AlBN/GaN conduction band offset $\Delta E_{\rm c}$ = 2.234 eV, and the measured thicknesses of the UID GaN cap layer of 2 nm, the AlBN barrier layer of 2.2 nm using the high-resolution transmission electron microscope images and X-ray reflectivity spectra. From the simulation, the relative dielectric constant of AlBN is estimated to be $\epsilon_{\rm r}^{\rm AlBN} \sim$16, compared to $\epsilon_{\rm r}^{\rm AlN}\sim$9 for AlN.  The sharp decrease in capacitance at negative voltage when the 2DEG depletes is consistent with the HEMT transfer characteristics seen in Figure \ref{fig4} (b).

%--------Figure 5---------------------------
%--------------------------
\begin{figure}[t!]
\centerline{\includegraphics[width=1\columnwidth]{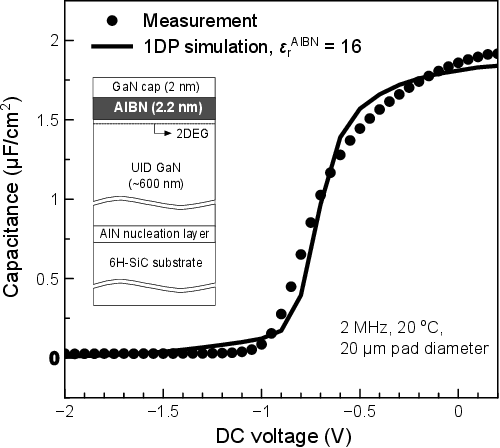}}
\caption{\label{fig5} Measured CV characteristics (points) of 2-nm GaN cap/2.2-nm AlBN/GaN-based metal-insulator-semiconductor (MIS) capacitor at 2 MHz, 20 $^\circ$C overlaid with 1D Schrödinger Poisson simulated CV characteristics indicating the dielectric constant of the AlBN barrier is $\epsilon_{\rm r}^{\rm AlBN} \sim$16.}
\end{figure}
%--------------------------
%--------Figure 5---------------------------

By performing the C-V measurements at 0.1, 0.5, 1, and 2 MHz frequencies (see supplementary information), we found the AlBN capacitance to be consistently higher than AlN, with the value decreasing slightly at high frequencies for both AlBN and AlN-based devices. This indicates a negligible effect of the space charge polarization and that the ionic and dipole polarization likely plays a role in enhancing the dielectric constant of AlBN. Theoretical studies indicated an enhancement in the dielectric constant of AlN with B incorporation due to increased ionicity from a change in local charge distribution, a decrease in structural stability, and/or softening of IR-active phonon modes.\cite{milne2023AlBN_BG_HK}  The higher dielectric constant seen in our MBE-grown AlBN thin films is consistent with the high-K dielectric nature reported in thick sputter-deposited AlBN films.\cite{Hayden2021AlBN} A slight difference in the observed dielectric constants between MBE-grown and sputter-deposited AlBN films could be due to different substrates and inherently different growth dynamics offered by MBE and sputtering. The epitaxial AlBN films in this work were grown on polar GaN semiconducting substrates, whereas the AlBN films reported by J. Hayden et al. were sputter deposited on nonpolar metallic tungsten layers.\cite{Hayden2021AlBN} Different strain conditions in the AlBN films grown on GaN by MBE vs. on metal by sputtering can impact the respective dielectric constants. The growth dynamics, resulting crystallinity, and chemical purity differ between MBE and sputter-deposition. Owing to the ultrahigh vacuum nature of MBE, elemental purity of sources, and minimum substrate damage from the low kinetic energies of evaporated molecular beams, MBE growth on epitaxial substrates yields films of high crystal quality, chemical purity, and controlled stoichiometry.\cite{nunn2021review} These factors can affect the local dipole moments and distributions, affecting the measured dielectric constant of the AlBN films grown by different methods and on different substrates. Regardless, the high-K nature of AlBN films over that of AlN films is unambiguous. The above results provide strong evidence that AlBN on GaN is an epitaxial high-K dielectric that can simultaneously provide polarization-induced 2DEGs and appropriate insulating properties to serve as the gate barrier of AlBN/GaN HEMTs.

Thus, boron-containing AlBN joins other transition or rare-earth metal high-K wurtzite nitrides such as AlScN, AlYN, AlLaN, and AlLuN.\cite{Hayden2021AlBN, fichtner2019alscn, wang2023YAlN, rowberg2021structural, utsumi1998BAlNpatent} The energy bandgap of AlN does not shrink substantially upon B substitution because the energy bandgaps of BN variants (5.96 eV for h-BN, 6.4 eV for c-BN, 6.84 eV for wurtzite-BN) are similar to AlN\cite{kudrawiec2020bandgap, shen2017band, milne2023AlBN_BG_HK, Hayden2021AlBN, suceava2023AlBNnonlinear, Savant2023AlBNMBEEMC}. In contrast, the energy bandgaps of the binary nitrides of Sc, Y, La, Lu are substantially smaller than AlN\cite{smith2001molecular, saha_electronic_2010, casamento2023review, Winiarski2019electronicstructure, winiarski2020crystal, larson_electronic_2007, leone2023AlYN}.  UV-visible spectroscopic measurements of thick AlBN MBE-grown films indicate a slight reduction of AlN bandgap from 6.1 eV to 6.0 eV with $\sim$7\% B incorporation, compared to 5.77 eV for $\sim$7\% Sc incorporation\cite{baeumler2019AlScNBG}.  Thus, the ultrawide bandgap of AlBN is expected to maintain a larger breakdown voltage than other hi-K nitride alternatives.

The route to incorporation of the non-transition group metal B in AlN differs from the group III transition metal Sc \cite{fichtner2019alscn, wang2023dawn} (and its variants such as Y\cite{wang2023YAlN, leone2023AlYN} or La\cite{rowberg2021structural, winiarski2020crystal}) in {\em four} essential ways: 1) the B-N chemical bonds do not involve d-orbitals - the competing crystal structure of BN is layered hexagonal with sp$^2$ bonds, and the less stable cubic and hexagonal wurtzite variants of BN have sp$^3$ bonds.\cite{kudrawiec2020bandgap}, 2) the $d_{33}$ piezoelectric coefficient does not increase (it slightly {\em decreases} in AlBN compared to AlN)\cite{suceava2023AlBNnonlinear, kudrawiec2020bandgap}, 3) since B atom is lighter than Al, the thermal conductivity of AlBN, though lower than AlN due to alloy scattering of phonons, is expected to be higher than for the heavier transition metal Sc, Y, or La alloys with AlN,\cite{alvarez_thermal_2023, jena2019new, utsumi1998BAlNpatent} and 4) AlBN is expected to be more resistant to oxidation compared to AlScN and other transition/rare-earth metal-based nitrides.\cite{casamento2020oxygen, greenwood_chemistry_1997}  Thus, AlBN has the potential to be an electronically, thermally, and chemically robust high-K dielectric suited for high temperature and harsh environment RF, power, and memory devices based on GaN.

In summary, the observations reported here about 1) the epitaxial growth of AlBN/GaN heterostructures, 2) the observation of polarization-induced 2DEG channels at the AlBN/GaN heterojunction, 3) the process compatibility to form HEMTs with ultrathin AlBN barriers, and 4) the observation of high-K dielectric properties of AlBN barrier layers offers significant hope for surpassing limitations imposed by AlGaN barriers in current GaN HEMT technology for power electronics and RF/mm-wave applications of this revolutionary semiconductor family.

\vspace{5mm}
\section*{Supplementary Material}
See the supplementary material for the 300 K Hall effect measurements map of a 3-inch diameter wafer with sheet resistivity, electron mobility, and 2DEG sheet density, with buffer leakage measurement and CV measurements on multiple devices with different top electrode diameters at different measurement frequencies confirming the high-K dielectric nature of the MBE AlBN film over the control AlN film.
\vspace{-2mm}

\vspace{5mm}
\section*{Author Contributions}
\vspace{-2mm}
Chandrashekhar Savant, Thai-Son Nguyen, and Kazuki Nomoto contributed equally to this work.

\begin{acknowledgments}
\vspace{-2mm}

This work was supported in part by the Ultra Materials for a Resilient Energy Grid (epitaxial growth, microscopy and characterization), an Energy Frontier Research Center funded by the U.S. Department of Energy, Office of Science, Basic Energy Sciences under Award $\#$DE-SC0021230, in part by SUPREME, one of seven centers in JUMP 2.0, a Semiconductor Research Corporation (SRC) program sponsored by DARPA (device fabrication) and in part by ARO Grant $\#$W911NF2220177 (device characterization).  This work used the Cornell Center for Materials Research (CCMR) Shared Facilities, which are supported by the NSF MRSEC program (DMR-1719875).  The authors acknowledge the use of the Cornell NanoScale Facility (CNF), a member of the National Nanotechnology Coordinated Infrastructure (NNCI), which is supported by the National Science Foundation (NSF Grant NNCI-2025233). 

\end{acknowledgments}
\vspace{-2mm}

\section*{Author Declarations}
\vspace{-5mm}
\subsection*{Conflict of Interest}
\vspace{-2mm}
The authors have no conflicts to disclose.

\section*{Data Availability}
\vspace{-2mm}
The data supporting this study's findings are available from the corresponding author upon reasonable request.

%\nocite{*}
\section*{References}
\vspace{-2mm}
\bibliography{Manuscript_final}% Produces the bibliography via BibTeX.

\end{document}